# An intracellular calcium frequency code model extended to the Riemann zeta function

Keith R Willison

*Institute of Chemical Biology, Department of Chemistry, Imperial College London, Molecular Sciences Research Hub, Wood Lane, London W12 0BZ, UK*

**ABSTRACT:** We have used the Nernst chemical potential treatment to couple the time domains of sodium and calcium ion channel opening and closing rates to the spatial domain of the diffusing waves of the travelling calcium ions inside single cells. The model is plausibly evolvable with respect to the origins of the molecular components and the scaling of the system from simple cells to neurons. The mixed chemical potentials are calculated by summing the concentrations or particle numbers of the two constituent ions which are pure numbers and thus dimensionless. Chemical potentials are true thermodynamic free Gibbs/Fermi energies and the forces acting on chemical flows are calculated from the natural logarithms of the particle numbers or their concentrations. The mixed chemical potential is converted to the time domain of an action potential by assuming that the injection of calcium ions accelerates depolarization in direct proportion to the amplitude of the total charge contribution of the calcium pulse. We assert that the natural logarithm of the real component ($\zeta_n$) of the imaginary term ($\zeta_n i$) of any Riemann zeta zero ($\frac{1}{2}+\zeta_n i$) corresponds to an instantaneous calcium potential ($Z_n$). In principle, in a physiologically plausible fashion, the first few thousand Riemann $\zeta$-zeros can be encoded on this chemical scale manifested as regulated step-changes in the amplitudes of naturally occurring calcium current transients. We show that pairs of $Z_n$ channels can form Dirac fences which encode the logarithmic spacings and summed amplitudes of any pair of Riemann zeros. Remarkably the beat frequencies of the pairings of the early frequency terms ($Z_n - Z_{n+1}$, $Z_n - Z_{n+2}$ ....) overlap the naturally occurring frequency modes ($\gamma$, $\delta$, $\theta$) in vertebrate brains. Action potential control of calcium transients is a process whereby neuronal systems construct precise step functions; actually Dirac distributions which also underpin the Riemann mathematics. The equation for the time domain in the biological model has a similar form to the Riemann zeta function on the half-plane and mimics analytical continuation on the complex plane. Once coupled to neurophysiological binding processes these transients may underpin calculation in eukaryotic nervous systems.

### Introduction

In most receptor-mediated signalling processes in biology the external input object, such as a photon or a hormone, is rapidly converted into a complex, transient chemical disturbance in the cell cytoplasm. A major problem in neuronal signalling is to understand how the information content of the nanoscale chemical perturbation is encoded into spike trains for onward transmission over great distances and how it is then decoded back into the spatial domain in the recipient cell.

---

*Email address: keith.willison@imperial.ac.uk

Calcium signalling is an ancient and ubiquitous information transfer system found in all eukaryotic cells and sodium channels evolved from calcium channels as eukaryotes developed excitable tissues and became larger in the Precambrian around 540 MYr ago (Liebeskind et al; 2011).

Although transient intracellular calcium concentrations are hundreds of times lower than the steady-state sodium ion concentration it is calcium, not sodium, that is the predominant signalling ion species in biology (Smedler and Uhlén, 2014). Our model treatment here directly couples chemical waves made from calcium ions (Laugh-lin, 2015) to classical Hodgkin-Huxley action potentials made from sodium ions and then generates a heterodyne signalling system. In classical signal processing terminology heterodynes are the consequence of mixing two signals at frequencies $f_1$ and $f_2$ to create two new signals, one at the sum $f_1 + f_2$ of the two frequencies, and the other at the difference $f_1 - f_2$. The mixer here is the hybrid action potential.

We use the Nernst chemical potential treatment to couple the time domains of sodium and calcium ion channel opening and closing rates to the spatial domain of the diffusing waves of the travelling calcium ions inside single cells. The hybrid chemical potentials are calculated by summing the charge contributions from the numbers of the two constituent ions which are pure numbers and thus dimensionless. Chemical potentials are true thermodynamic free Gibbs/Fermi energies and the forces acting on chemical flows are calculated from the natural logarithms of the particle numbers or their concentrations. The mixed chemical potential is converted to the time domain of an action potential by assuming that the injection of calcium ions accelerates depolarization in direct proportion to the amplitude of the total charge contribution of the calcium pulse.

In principle, in a physiologically plausible fashion, the first few thousand Riemann $\zeta$-zeros could be encoded on this chemical scale manifested as regulated step-changes in the amplitudes of naturally occurring calcium current transients. We show that pairs of $Z_n$ channels could encode the spacings and summed amplitudes of any pair of Riemann zeros and can be described as Dirac fences with Shah function properties where the Fourier transform of the signal is itself. Remarkably the beat frequencies of the frequency terms of hundreds of the early pairings ($Z_n$ - $Z_{n+1}$) overlap the naturally occurring frequency modes ($\gamma$, $\delta$, $\theta$) in vertebrate brains. Action potential control of calcium transients, via actin-regulated channels, is a natural process whereby neuronal systems construct precise step functions which are actually Dirac distributions that also underpin the Riemann mathematics. It is possible that these distributions have been adapted further by neurophysiological binding processes to permit calculation in eukaryotic nervous systems and ultimately to encode number representation in human brains.

**Chemical potential treatment**

Chemical potential is an intensive quantity like temperature and always refers to a specific substance even though that substance may be a component of a mixture (Job and Herrmann, 2006). Almost invariably, chemical potentials are manipulated as changes between two states rather than their absolute values. Here we use chemical potential analysis to calculate the instantaneous forces delivered by flows of sodium and calcium ions as they cross cell membranes, through specific, selective ion channels, by travelling down their respective chemical gradients from outside to inside cells.

Equation 1: The chemical potential equation quantifying forces acting on chemical flows is

$$\mu_i = \mu_i^0 + RT \cdot \ln n_i$$

Because the chemical potential ($\mu$) is a measure of how much the free energy of a system changes upon addition to or removal of a number of particles ($n_i$) from the reference state, $\mu_i^0$, the units are those of an energy, a Gibbs energy, $G$. In our treatment the concentrations of ions are used for calculations and therefore the chemical potential is an energy density; the partial derivatives of which can be thought of as generalized forces.

During the rising phase of a classical Hodgkin-Huxley action potential, sodium ($Na^+$) ion fluxes can be defined using the Nernst equilibrium potential equation which has the same form as Eq1.

Equation 2: Nernst equilibrium potential equation is

$$E_i = \frac{RT}{z_i F} \ln \left( \frac{c_i^{out}}{c_i^{in}} \right)$$

For $Na^+$, the concentration gradient values in millimoles are $C^{in}$ = 10mM, $C^{out}$ =140mM. The potential energy is $E_{Na+}$ = +71mV. Nernst potentials are calculated at $T$=310°K and constant pressure.

Equation 3: Summing the sodium and calcium potentials

$$\mu_i = \mu_i^0 + RT \cdot \ln Na^+ + \ln Ca^{2+}$$

This paper develops the idea of a two component action potential composed of a mixture of sodium and calcium ions whose overall total charge remains constant but whose firing time is advanced by the addition of "extra" calcium ions during the rising phase. The total chemical potential of the hybrid action potential is the sum of the natural logarithms of each of the two ion currents.

For $Ca^{2+}$ the concentration gradient values are $C^{in}$ = 0.0001mM, $C^{out}$ =1.2mM. The potential energy is $E_{Ca2+}$ = +125mV.

Equation 4: Adjustment for the fugacity of calcium

We note that calcium ($Ca^{2+}$) is a doubly positive ion compared to the singly charged sodium ion ($Na^+$). For the parental sodium-only action potential containing N particles and total charge $N^+$ the number of particles in the hybrid action potential with the same charge = $N\,Na^+ - ½\,Ca^{2+}$ charge contribution. Every calcium ion replaces two sodium ions which yields a linear relationship at all scales.

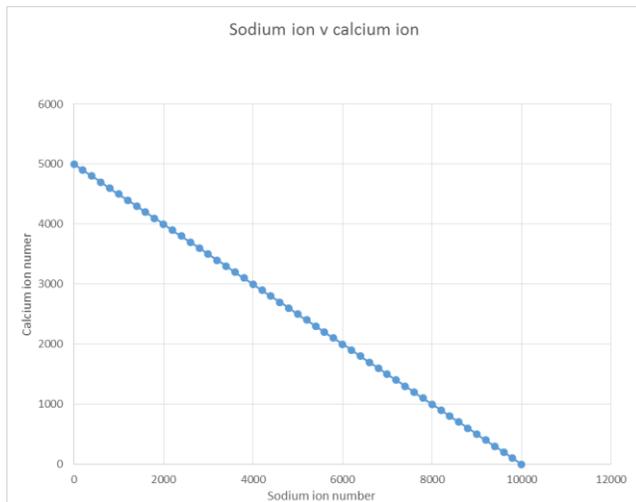

Figure 1.
N $Ca^{2+}$/ N $Na^+$ = tan θ = ½
The total charge amplitude is always set fixed for any particular species of spike train and has a value of 1 for frequency calculations (Equation 5).

The 'conductance' of the system is a complicated property to calculate because sodium and calcium cannot be in equilibrium at the same time and their absolute concentrations are ~100 fold different. Furthermore the protein ion channels that transport them in and out of cells are different, and other ionic flows are also in play.

To simplify the analysis we are going to consider the chemical potential of calcium as an instantaneous injection into the rising action potential. It is known from experimental measurements that the relative contribution of calcium ions to an action potential is only a few per cent of the dominant sodium current. We do not want to get into too much mechanistic development of this aspect of our model here but it is important to provide an estimate the potential dynamic range of the calcium current. Consider a dendritic bouton of volume 0.1pl receiving a bolus of calcium ions through a channel which increases local [$Ca^{2+}$free] by 1μM: 1 mole/litre = 6 x $10^{23}$ molecules and 1μM/0.1pl = 6x$10^4$ molecules. We have used a minimum estimate for the volume of a dendritic bouton. Dendrites are the computation centres in single neurons (London and Hausser, 2005). Obviously if the calcium channel component is multiplexed into the action potential integral then the dynamic range will increase. The time resolution of each calcium current injection is comfortably in the low millisecond range since protein ion channels can permit ion flows of ~$10^7$ sec$^{-1}$. It is an essential requirement that a neuronal calculation system is fast and action potentials satisfy this condition.

**Calcium diffusion for a two channel model**

Let us first consider two identical calcium channels, closely located in a membrane, which open and close at the same time; in phase (Figure 2). The two pulses of calcium ions will diffuse into the cytoplasm in a semi-circular wave and the location where the wave-fronts meet will constitute the maximum amplitude of the calcium signal. The absolute value of the maximum amplitude will be the sum of the two signals minus some constant terms which depend upon the rate of diffusion and dispersion of the calcium ions into the half-sphere. The physical location of the wave-front peak is estimated to be of the order of a micron or less (calculation not shown). If we consider two non-identical channels the wave front peak location will be steered towards the 'weaker' channel; the channel which opens for less time and therefore transmits fewer calcium ions. The diffusion coefficient of calcium is 2000 m²s⁻¹.

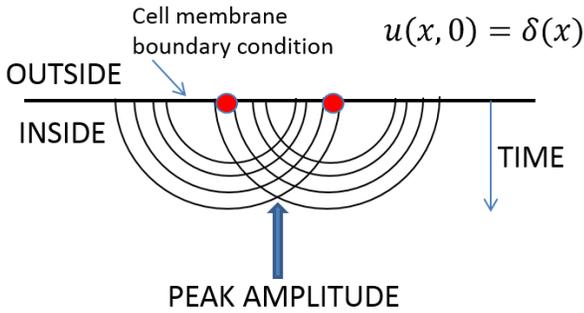

● indicates a calcium channel in the plane of the membrane permitting flow of Ca²⁺ ions into cells

Figure 2.
Two identical calcium channels (red circles), closely located in a membrane, open and close together The two pulses of calcium ions diffuse into the cytoplasm in semi-circular waves and the location where the wave-fronts meet constitutes the maximum amplitude of the paired signal (blue filled arrow).

***In vivo* calcium transients overlap Riemann zeros**

We assert that the natural logarithm of the real component ($\zeta_n$) of the imaginary term ($\zeta_n i$) of any Riemann zeta zero (½+$\zeta_n i$) corresponds to an instantaneous calcium potential ($Z_n$).

Assertion:
Ln $\zeta_n$ is equivalent to ln ($Ca^{2+}_{out}/Ca^{2+}_{in}$) = $Z_n$

Let us compare the logarithms of the imaginary terms of the $\zeta$-zeros with the amplitudes of the calcium currents (Appendix 1). The concentrations of calcium outside neurons, $[Ca^{2+}]_o$, compared to free-calcium inside cells $[Ca^{2+}]_i$ at the Nernst equilibrium potential for calcium (+125mV) are between 2.5-1.2mM versus 0.0001mM respectively. Calcium enters cells through specific ion channels, gated by active F-actin processes, and contributes to a rising Na⁺ action potential and the free-calcium concentration rises into the micromolar range. Therefore we can calculate the range of values for the calcium current flows as ln ($[Ca^{2+}]_o / [Ca^{2+}]_i$).

Figure 3 shows a plot of selected points on this range compared to the natural logarithms of the first 500 real numbers of the imaginary terms of (1/2 + i$\zeta$). It is striking how the two distributions overlap in the steepest phase of the exponential distribution of the calcium concentration range and continue in the limit when the free calcium concentration is in the nanomolar range (between 0 and 1 on the plot).

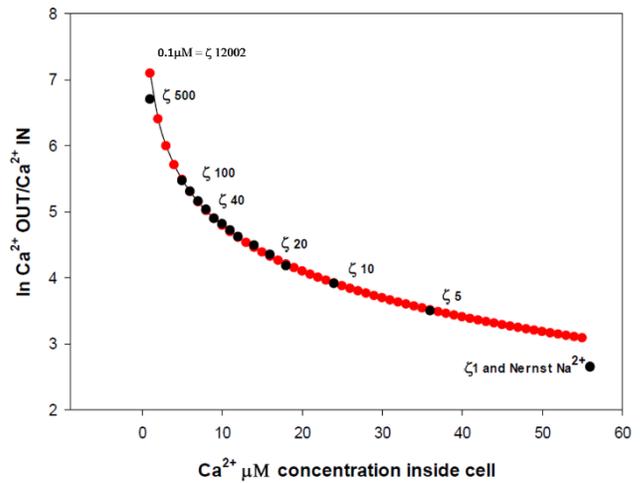

Figure 3.
Plot of ln calcium transients (1-56μM) and co-mapping of $\zeta$-zeros 1-500. The red trace shows ln Ca²⁺OUT/ Ca²⁺IN for 1-56μM internal [calcium] in 1μM steps ([Ca2⁺OUT] is constant at 1200μM since this external pool is large both in concentration and volume; the inflow of calcium does not perturb the pool). The distribution is actually smooth at this scale since each μM step contains a minimum of 6 x 10⁴ ions. The black dots on the red curve show ln of the imaginary term of selected Riemann $\zeta$-zeros up to $\zeta$500. Interestingly ln $\zeta$1 and the Nernst equilibrium potential for sodium are similar: Na⁺ = ln 140/10 and $\zeta$1 = 14.13473 is equivalent to [Ca⁺IN] = 84.9 (Note: x-axis is truncated at 60). The numbers generated during this mapping are shown in Table 1 for the first 25 Riemann $\zeta$-zeros.

**Conversion of chemical potential equation to the time domain**

The following equation converts an instantaneous calcium potential (Ca-in), expressed in units of micromoles, to a spike train frequency carried by a specific sodium action potential type.

Equation 5: Converting instantaneous potential to frequency

Na$^+$ channel rate (sec$^{-1}$). [Na$^+$-Resting (μM)].
$$\frac{1}{([\text{Na}-\text{Resting }(\mu M)] - 2.[\text{Ca}-\text{in }(\mu M)])}$$

Units are seconds$^{-1}$

Appendix 1 shows the detailed workflow and numbers behind equation 5. Table 1 shows the values for converting the first 25 zeros to the chemical domain and then to the frequency domain. There is a direct reciprocal relationship between the frequencies and spatial domains of the ion channel chemical potentials analogous to the properties of Shah functions (*III*) where the Fourier transform of a Shah function is itself. The spatial domain diffusive-ion structure can be considered as a Dirac fence in two dimensions. The chemical component therefore contains the property of convolution, ($C(x)$) of $F_1(x)$ with $F_2(x)$ where the Fourier pair is the product of $\phi_1(p)$ and $\phi_2(p)$, because logarithms are being summed in the chemical potential equation.

**Beat frequencies and transformation to and from physical space**

In acoustics a beat is caused by interference between two sounds of slightly different frequencies. This periodic variation in volume occurs at a rate equal to the difference between the two frequencies. The derivation of the beat frequencies of constructive and destructive interference of sine waves requires only straightforward trigonometry (not shown).

Equation 6: $\quad f_{\text{beat}} = f_1 - f_2$

Equation 7: Beat frequency calculation for pairing chemical zeros $Zn$

$$f_{\text{beat}} = \frac{1}{1 - \ln\frac{Ca\ out}{Ca\ in} Zn+1} - \frac{1}{1 - \ln\frac{Ca\ out}{Ca\ in} Zn}$$

$Zn$ is the largest zero in any particular series

Beat frequencies occur in any vibrating system and in this case the beats are nodes of increased calcium concentration adding constructively only, since we cannot have negative calcium ion concentrations.

Figure 4 shows a plot of the beat frequencies and the calcium amplitudes of the first four Riemann zeta zero terms, z1, z2, z3 and z4 coupled to a 1 kHz sodium channel (Table 1) and paired with the subsequent zeros up to z100. Each series forms a straight line in this space and all the lines in this representation are parallel. Remarkably the beat frequencies range from ~1Hz to 17Hz for the first 100 zeros and overlap the natural frequency range of vertebrate brain oscillations (Buzsaki and Draguhn 2004).

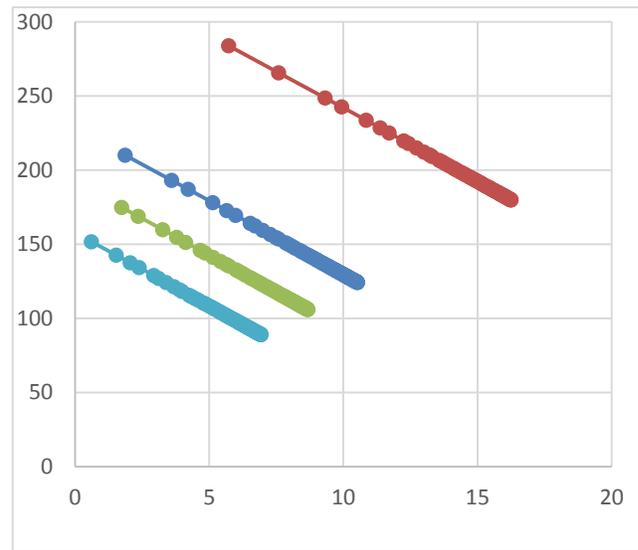

Figure 4.

Beat frequencies of interacting calcium amplitudes. Red line is a plot of the 99 beat frequencies of z1-z2 ……z1-z100. Light blue line is a plot of the 98 beat frequencies of z2-z3 ……z2-z100. Green line is a plot of the 97 beat frequencies of z3-z4 ……z4-z100. Turquoise line is a plot of the 96 beat frequencies of z4-z5 ……z4-z100. The y-axis shows the physical amplitude of the mixed signals which are the addition of the values for the calcium transient shown in Table 1 (Fugacity column).

Figure 5 shows the beat frequency profiles of pairs of calcium transients with repeated, random and Riemann zero spacings over the chemical space. Each descending series forms a line in beat frequency space which decreases in amplitude as the calcium transient terms become smaller. For a given spacing mechanism all the lines are parallel which is most easily viewed for the evenly divided space (Fig 5A) because a simple plane is observed.

A single random division run yields a distorted plane because the gap sizes vary but if the mean of many random runs were taken a planar surface would be observed. The Riemann spacings yield a beautiful manifold which runs towards the beat frequency limit of zero as the terms reduce in amplitude. We note that pairs of channels with identical frequencies (Fig 2) will have a chemical amplitude but no beat frequency value; $f_{beat} = f_1 - f_1 = 0$. These identical pairs are summed logarithms and it is interesting to note that the Riemann function has support at the squares of the primes. Although the beat frequency would be physically it is a secondary property of the behaviour of the chemical system.

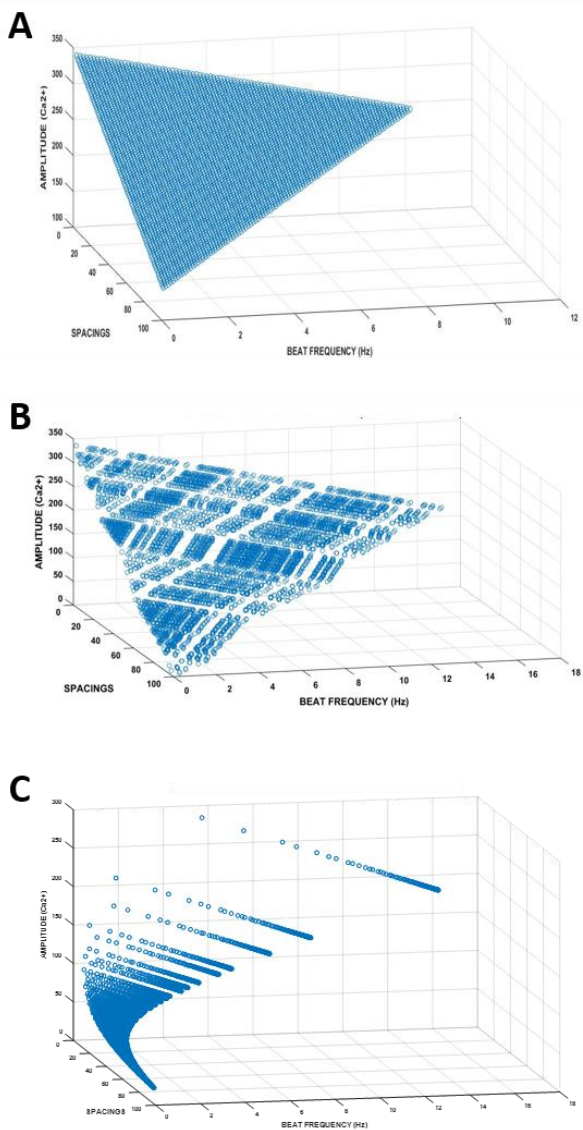

Figure 5

Panels A, B, C show beat frequencies over the range of the calcium gradient ($Ca^{2+}$ in) shown in column 6 (Table 1). Fig 5A divides the calcium gradient evenly into 99 segments, producing a plane in the state space. Fig 5B shows 100 random real numbers (Matlab – rand function), aligned in descending order and dividing the space into 99 segments. Figure 5C shows the state space divided up into 99 segments using the first 100 Riemann zeros. The z-axis shows the value of the chemical amplitude of the pairs.

**Discussion**

This paper presents a new way of considering the relationship between chemical potentials and action potential generation in cell signalling. The values and behaviours of the chemical ion components are known with high confidence. An important aspect has been to minimize the biological detail in order to generate the mathematical description. The outcome is a straightforward state space which includes both the chemical and time domains of the hybrid system. The scale of the chemical space is highly restricted because of the low dynamic range of the calcium gradient which exists across the membranes of eukaryotic cells. The scale is logarithmic and this fact has implications for understanding the origin of the logarithmic properties of sensation scales; the Weber-Fechner Law. Furthermore, despite the restricted scale, the model of the system yields a range of beat frequencies which exactly lie on the conventional brain-wave scale (eg $\gamma$, $\delta$, $\theta$) which is itself a logarithmic scale found in all vertebrate brains (Buzsaki and Draguhn, 2004).

We show that the calcium scale can be divided up in several ways (Figure 5) which is unsurprising since it is a real number line but we have focused on dividing the space guided by the properties of the Riemann zeta function zeros. One intention of our work was to think about effective methods to divide up physical space inside cells where order disruption by diffusion and thermalization processes abounds. It is one of the great innovations of eukaryotes that although their cells are much larger than bacterial species and are only enclosed by single lipid bilayer membranes they are able to organize and sub-divide their cytoplasm effectively and reproducibly. It was thinking about divisibility and segmentation that caused us to connect our model to discontinuous functions and the Riemann zeta function in particular.

**ζ-zeros and the distribution of prime numbers**

The number of prime numbers less than or equal to ξ, π(ξ), is well approximated by simple logarithmic integrals. Prime number theorem (PNT) asserts that π(ξ) is asymptotic to ξ /ln ξ and Gauss's integral is expressed as the reciprocal of the logarithm = $\int_2^t \frac{1}{\log t} dt$. This integral is called the logarithmic integral of ξ, denoted as $Li(x)$. The great insight of Riemann was the enhancement of the $Li(x)$ integral with an infinite series of integrals,

$$(x) = 1 + \sum_{m=1}^{\infty} \frac{(\log x)^m}{m! m \zeta(m+1)\infty}.$$

In this equation ζ is the Riemann ζ -function $\sum_n n^{-s} = \prod_p (1-p^{-s})^{-1}$.

Note that variable nomenclature is used for the real part of the imaginary term of the zero (ζ, σ and *t*).

Essentially the ζ -function is a correction term which improves the approximation to π(ξ). We note that PNT is intimately related to natural logs and no other flavours of logarithms (Montgomery and Wagon, 2006) as is the Nernst equilibrium potential equation. Interestingly the amplitude of the correction terms relative to $Li(x)$ and the calcium amplitudes relative to sodium overlap; the low lying zeros contribute a maximum of ~ 2% to $Li(x)$ and the biological system has a maximum effect on the sodium channel firing rate which is estimated at 1.7 % for a single injection of current (first term in Table 1 – 84.9). We remark that our equation for converting the Riemann zeros to frequencies has a similar form to the ζ -function despite operating in different mathematical spaces and the original zeta function of Euler over the reals.

**The Riemannium**

Polya and Hilbert famously suggested that there is a naturally occurring Hermitian operator whose eigenvalues are the zeros of ζ(1/2+*it*) and are therefore real and that the heights of the zeros correspond to the frequencies of an unknown vibrating system. The mathematical search for the vibrating system has predicted the algebraic properties of the so called Riemannium (Bohigas, Lebœuf and Sánchez, 2001; Leboeuf, Monastra and Bohigas, 2001) and computational investigations have made discoveries about the properties and behaviours of the GUE system by examining the amplitudes of the zeros and the nearest-neighbour spacings between the zeros (Odlyzko, 1987). The fields of classical and quantum physics have also been the landscape for the search for the operator because Riemann zeta functions and several other types of related zeta functions and Dirichlet L-series all seem very effective in describing natural physical systems (Lapidus, 2008; Schumayer and Hutchinson, 2011). Even the realm of quantum chaology has been linked to the Riemann zeros (Berry and Keating, 2013). Dyson proposed an inverse approach to this problem (Dyson, 2010) stating that if the Riemann hypothesis is true, then the zeros of the zeta-function form a one dimensional quasi-crystal according to the definition and constitute a distribution of point masses on a straight line, and their Fourier transform is likewise a distribution of point masses, one at each of the logarithms of ordinary prime numbers and prime-power numbers. Our model resembles Dyson's quasi-crystal idea but differs considerably in that it also contains its own intrinsic frequency domain; the ascending beat frequencies being aligned along a one-dimensional line of points containing all the pairs of zeros in descending order (Figure 5). The biological line is chemically constrained but mathematically speaking it could be infinite. In both systems precision is expensive because more correction terms are required as the systems grow in size. In the slowly ascending Riemann series the integrals on ℂ slowly become increasingly huge in order to effect correction higher up the number line (Review: Conrey, 2003). Biologically speaking, correction occurs as the individual calcium pulses become smaller and this is because enhancement of the precision of the chemical system would require more components and more energy. We note that Riemann spacings are a maximum entropy solution, S=w ln w, to dividing number lines (Katz and Sarnak, 1999 ) and it is likely that evolution too found the maximum entropy solution to dividing up space and time. Presumably this MaxEnt solution (Jaynes, 2003) has reached its apogee in the Bayesian calculation centres in human brains. Riemann mathematics involves analytical continuation on the complex plane and we note that the points on the beat frequency rays have unique amplitudes and therefore knowing one value allows continuation of the series. This property suggests how single action potential spikes could transfer information via the spatial domain in addition to the time-gap between a pair of spikes (Rieke et al; 1999). A further similarity is that the Riemann function is supported at the prime powers and in our state space the pairs of identical channels have beat frequencies of zero but amplitudes of twice the log of the calcium potential.

To our knowledge no-one has tried to build a biological model for the operator and it is our goal to develop it further. The Riemann hypothesis has a mythical attraction because it addresses an essential enigma; where does our mathematics come from? We assume that mathematics is generated by standard activities in our brains and we are building an operator using known molecules and well understood biological phenomena and estimations of the numbers and kinetics thereof. We are not invoking new physics to explain an emergent biological phenomenon. Our model scales with system size and therefore incorporates the concept of evolvability; mathematical ability cannot have originated *de novo* in early humanoids, rather it must have built upon pre-existing calculation systems already in use in vertebrates and probably invertebrates too.

**Acknowledgments**

**References**

Berry, M. V and Keating, J. P. (1999) 'The Riemann zeros and eigenvalue asymptotoics', SIAM Review, 41, pp. 236-266.

Bohigas, O., Lebœuf, P. and Sánchez, M. J. (2001) 'Spectral spacing correlations for chaotic and disordered systems', *Foundations of Physics*, 31(3), pp. 489–517. doi: 10.1023/A:1017569612944.

Conrey, J.B. (2003) The Riemann Hypothesis. Notices AMS, 50(3):341-353.

Draguhn, A. and Buzsaki, G. (2004) 'Neuronal Oscillations in Cortical Networks', *Science*, 304(June), pp. 1926–1929.

Dyson, F. (2010) 'Birds and frogs in mathematics and physics', *Uspekhi Fizicheskih Nauk*, 180(8), p. 859. doi: 10.3367/UFNr.0180.201008f.0859.

Jaynes, E. (2003) *Probability Theory: The Logic of Science*. 5th printing. Edited by G. Bretthorst, Larry. Cambridge: Cambridge University Press.

Job, G. and Herrmann, F. (2006) 'Chemical potential—a quantity in search of recognition', *European Journal of Physics*, 27(2), pp. 353–371. doi: 10.1088/0143-0807/27/2/018.

Katz, N. M. and Sarnak, P. (1999) Zeros of zeta functions and symmetry.*Bull. American. Math.Soc, 36(1), pp1-26.*

Lapidus, M. L. (2008) *In Search of the Riemann Zeros*. American Mathematical Society.

Leboeuf, P., Monastra, a. G. and Bohigas, O. (2001) 'The Riemannium', p. 10. doi: 10.1070/RD2001v006n02ABEH000170.

Laughlin, R.B. (2015) Critical waves and the length problem of biology. PNAS, 112(33): 10371-10376.

Liebeskind, B.A. Hillis, D.M. Zakon, H.H. (2011) Evolution of sodium channels predates the origin of the nervous systems in animals. PNAS, 108(22): 9154-9159.

London, M. and Hausser, M. (2005) Dendritic Computation. Ann Rev Neurosci 28:503-532.

Montgomery, H. L. and Wagon, S. (2006) 'A heuristic for the prime number theorem', *Mathematical Intelligencer*, 28(3), pp. 6–9. doi: 10.1007/BF02986877.

Odlyzko, A. M. (1987), "On the distribution of spacings between zeros of the zeta function", Mathematics of Computation, 48 (177): 273–308, doi:10.2307/2007890.

Rieke, F, Warland, D, de Ruyter van Steveninck, R and Bialek, W. (1999) Spikes: exploring the neural code. MIT Press.

Schumayer, D. and Hutchinson, D. A. W. (2011) 'Colloquium: Physics of the Riemann hypothesis', *Reviews of Modern Physics*, 83(2), pp. 307–330. doi: 10.1103/RevModPhys.83.307.

Smedler, E. and Uhlén, P. (2014) 'Frequency decoding of calcium oscillations', *Biochimica et Biophysica Acta - General Subjects*. The Authors, 1840(3), pp. 964–969. doi: 10.1016/j.bbagen.2013.11.015.

Appendix 1

Mathematical workflow for calculating instantaneous calcium transients to beat frequencies

**WORKFLOW**

1. The calcium ion gradient *in vivo* is; $Ca^{2+}_{out}$ = 1200 μM and $Ca^{2+}_{in}$ = 0.1 μM (effectively zero).
2. The sodium ion gradient *in vivo* is; $Na^{1+}_{out}$ = 140000 μM and $Na^{1+}_{in}$ = 10000 μM.
3. In the model a single injection of calcium ions is deposited during the evolution of a single sodium action potential. Since the amplitude of the calcium ion component is around only 1% of the total amplitude of the action potential we can ignore time and de-phasing as variables during the calcium influx process and assume an instantaneous deposition process[1]. Nevertheless, because the absolute amplitude of the action potential, and therefore firing timing, is constant for any particular rapid-fire spike train, calcium ion addition to an individual action potential will slightly accelerate the discharge rate.
4. Calculations may be performed using moles or particle numbers. Using moles avoids taking account of the absolute size of the action potentials and the system size. We take the selected value[2] for a single calcium influx ($Z_n$) in micromoles and divide 1200 μM by $Z_n$. The natural logarithm of this ratio yields the chemical potential energy distribution[3]
5. Multiply 1200 / $Z_n$ by 2 because calcium is a doubly charged positive ion whereas sodium is a single charged positive ion.
6. The sodium ion concentration of an action potential at the time of firing is constant in this model and in these initial calculations is set at 10000 μM, the resting state concentration inside cells[4].
7. Therefore a mole of sodium action potentials will contain 10000 μM of sodium charges but a mole of hybrid action potentials will consist of 10000 – (2 x 1200 / $Z_n$) moles of sodium charges and 2 x 1200 / $Z_n$ moles of calcium charges.
8. The hybrid action potential will fill up and discharge earlier in proportion to the 'extra' calcium ions deposited. The acceleration of the firing time is the reciprocal value = 1 / [10000 – (2 x 1200 / $Z_n$)].
9. Multiply 1 / [10000 – (2 x 1200 / $Z_n$)] by 10000 to set the value of the sodium only action potential *amplitude to 1*.[5]
10. Select a frequency for a train of spikes. For example if this is to be set at 1 kHz then further multiply 1000 to convert to Hz.

---

[1] Since our model is as minimal as possible we assume an action potential generated by multiple sodium channels and a single calcium channel. Suppose that the mean action potential firing rate is 1000Hz and the mean calcium channel rate is 1000Hz, each with some variance around the mean, then the activities of the two channel systems are invariably in phase and co-incident.

[2] See Table 1 column 4. We note that the maximum value of an instantaneous influx is unlikely ever to be greater than 100 μM even with multiple calcium channel injections into one mixed action potential.

[3] Energy distribution because chemical concentrations are being manipulated. Calculations using particle numbers yield pure energies.

[4] During a neuronal action potential large changes take place in the membrane potential but the actual concentrations of the positive ions $Na^+$ and $K^+$ do not change. Approximately 2 x $10^6$ $Na^+$ ions enter the cell during a single spike phase representing ~0.06% of the total number per cell.

[5] This transforms the spikes into Dirac distributions and scales the system with respect to the channel numbers and phasing (see footnote 1).

**Equation to convert an instantaneous calcium potential ($Z_n$) to a spike train frequency carried by a specific sodium action potential type**

Sodium channel rate (sec$^{-1}$). [Na$^+$-Resting (μM)]. $\frac{1}{([\text{Na}-\text{Resting (μM)}]-2.[\text{Ca}-\text{in (μM)}])}$

**Units are seconds$^{-1}$**

$$F_{Z_n} = 10^3 \cdot 10^4 \cdot \frac{1}{(10^4 - 2 \times Zn)}$$

**In this example the spike train frequency is 1kHz**

**STATEMENT: Let the natural logarithm of the real component of the imaginary term of the Riemann zeta zero ($1/2+\zeta_n$ i) correspond to an instantaneous calcium potential.**

**Ln $\zeta$n is equivalent to ln ($Ca^{2+}_{out}/Ca^{2+}_{in}$)**

**Table 1**

| Riemann zero | Log zero | Ca2+ out | Ca2+ in | Log Ca2+ out/in | Fugacity | Na+ - Ca2+ |
|---|---|---|---|---|---|---|
| 14.13472514 | 2.648634546 | 1200 | 84.9 | 2.648602742 | 169.8 | 9830.2 |
| 21.02203964 | 3.045571394 | 1200 | 57.1 | 3.045272719 | 114.2 | 9885.8 |
| 25.01085758 | 3.219310034 | 1200 | 48 | 3.218875825 | 96 | 9904 |
| 30.42487613 | 3.415260567 | 1200 | 39.45 | 3.415042788 | 78.9 | 9921.1 |
| 32.93506159 | 3.494537792 | 1200 | 36.5 | 3.492764575 | 73 | 9927 |
| 37.58617816 | 3.626636381 | 1200 | 32 | 3.624340933 | 64 | 9936 |
| 40.91871901 | 3.711587636 | 1200 | 29.17 | 3.716936052 | 58.34 | 9941.66 |
| 43.32707328 | 3.768777689 | 1200 | 27 | 3.79423997 | 54 | 9946 |
| 48.00515088 | 3.871308315 | 1200 | 25.9 | 3.835833867 | 51.8 | 9948.2 |
| 49.77383248 | 3.907489394 | 1200 | 24 | 3.912023005 | 48 | 9952 |
| 52.97032148 | 3.969731785 | 1200 | 22.9 | 3.958939925 | 45.8 | 9954.2 |
| 56.4462477 | 4.033288817 | 1200 | 21.9 | 4.003590199 | 43.8 | 9956.2 |
| 59.347044 | 4.083402314 | 1200 | 20.1 | 4.089357021 | 40.2 | 9959.8 |
| 60.83177852 | 4.108112326 | 1200 | 19.8 | 4.104394898 | 39.6 | 9960.4 |
| 65.11254405 | 4.17611722 | 1200 | 18.4 | 4.177726171 | 36.8 | 9963.2 |
| 67.07981053 | 4.205883112 | 1200 | 17.9 | 4.205276123 | 35.8 | 9964.2 |
| 69.54640171 | 4.24199418 | 1200 | 17.2 | 4.245167452 | 34.4 | 9965.6 |
| 72.06715767 | 4.27759843 | 1200 | 16.7 | 4.274668116 | 33.4 | 9966.6 |
| 75.7046907 | 4.326840123 | 1200 | 16 | 4.317488114 | 32 | 9968 |
| 77.14484007 | 4.345684695 | 1200 | 15.6 | 4.342805922 | 31.2 | 9968.8 |
| 79.33737502 | 4.373709329 | 1200 | 15.1 | 4.375382092 | 30.2 | 9969.8 |
| 82.91038085 | 4.417760276 | 1200 | 14.7 | 4.402229342 | 29.4 | 9970.6 |
| 84.73549298 | 4.439534557 | 1200 | 14.4 | 4.422848629 | 28.8 | 9971.2 |
| 87.42527461 | 4.470784424 | 1200 | 13.7 | 4.472681003 | 27.4 | 9972.6 |
| 88.80911121 | 4.486489248 | 1200 | 13.5 | 4.48738715 | 27 | 9973 |

Table 1 shows the numerics of the conversion of the real part of the imaginary term for the first 25 Riemann zeros. The natural logarithm of the zero is located on the calcium chemical potential gradient (here to 2 significant figures). The effective instantaneous concentration (here given in units of micromoles) is then doubled to take account of the double positive charge of calcium ions compared to sodium ions (Fugacity). This number is then subtracted from the effective sodium ion concentration inside cells (10mM i.e. 10000 micromolar). In the reciprocal space of the time domain this term is the adjustment factor of the sodium channel spiking frequency. It is evident that, physiologically speaking, a large change over the calcium gradient (57.1-13.5 = 43.6 micromoles) has a small effect in the hybrid channel because the sodium ion gradient is ~ 1000x larger.